\documentstyle[preprint,aps,epsfig,floats]{revtex}

\tightenlines
\draft
\begin{document}
\date{\today}
\preprint{\vbox{\hbox{UA/NPPS-16-98}\hbox{RUB-TPII-18/98}}}
\title{Off-mass-shell Sudakov-like suppression factor for the
       fermionic four-point function in QCD \\}
\author{A. I. Karanikas,${}^{1}$\thanks{E-mail:
        akaranik@cc.uoa.gr}
        C. N. Ktorides,${}^{1}$\thanks{E-mail:
        cktorid@cc.uoa.gr}
        N. G. Stefanis,${}^{2}$\thanks{E-mail:
        stefanis@tp2.ruhr-uni-bochum.de}
        and
        S. M. H. Wong${}^{1}$\thanks{Present address:
        Physics Department, University of Minnesota, Minneapolis,
        MN 55453, USA}
        \thanks{E-mail: swong@mnhepw.hep.umn.edu}
        }
\address{${}^{1}$ University of Athens,                     \\
                  Department of Physics,                    \\
                  Nuclear and Particle Physics Section,     \\
                  Panepistimiopolis,                        \\
                  GR-15771 Athens, Greece                   \\
                  [0.3cm]
         ${}^{2}$ Institut f\"ur Theoretische Physik II     \\
                  Ruhr-Universit\"at Bochum                 \\
                  D-44780 Bochum, Germany                   \\
                  [0.3cm]
         }
\maketitle
\newpage
\begin{abstract}
We consider a four-point process, associated with a wide-angle
elastic scattering of two {\it off-mass-shell} spin-1/2 matter
particles, in a non-abelian gauge theory. On the basis of a worldline
approach, which reverts the functional to a path-integral description
of the system, we factorize an {\it eikonal} (``soft'') subsector of
the full theory and calculate the Sudakov-like suppression factor for
the four-point function as a {\it whole}, once we have extracted the
associated anomalous dimensions and taken into account the
renormalization-group controlled evolution.
\end{abstract}
\pacs{11.10.Hi,11.10.Jj,12.38.Aw,12.38.Cy}
\newpage   
\input amssym.def
\input amssym.tex
{\bf 1.} Next to confinement, factorization defines the single most
important issue associated with the theoretical confrontation of QCD.
In particular, a clean passage from partonic to hadronic modes of
description entails the formation of infrared-safe quantities, wherein
long-distance (nonperturbative) contributions appear in a factorized
form.
A prime example of isolating long-distance behavior in the context of
a purely field theoretical calculation, involving quarks and gluons, is
provided by the Sudakov form factor (see, for example,
\cite{DDT80,Col89,Mue81} for reviews).
Recently, two of us have established \cite{KK98} an analogous
mode of behavior which refers to a four-point process, pertaining to
on-mass-shell elastic quark-quark scattering at a fixed angle, viewing
this process not in terms of Feynman graph lines or as part of an
operator product expansion (OPE) based treatment but rather as a {\it
whole}.

Our approach to factorization relies on the so-called worldline
casting of a non-abelian gauge field theory with spin-1/2 matter fields
\cite{KK92}, which results from the reformulation
$
 \int_{}^{}{\cal D}\bar{\psi}(x) \, {\cal D}\psi (x) \,
 {\rm e}^{S\left[ \bar{\psi}(x), \psi (x), A_\mu (x) \right]}
 \ldots
\longrightarrow
 \int
 {\cal D}x(\tau) \, {\cal D}p(\tau) \,
 {\rm e}^{S\left[ x(\tau), p(\tau), A_\mu\left( x(\tau) \right) \right]}
\ldots
$,
taking us from a functional to a path-integral description of the
system.
In particular, working explicitly in the Feynman gauge, we have
argued \cite{KKS95,GKKS97} that contributions to the path integral,
which take into consideration only those paths that are straight lines
almost everywhere (allowing, therefore, for the presence of cusps) and
with the Dirac determinant set to unity, factorize in a most natural
manner.
In physical terms, the above specifications enable the isolation of a
subsector of the full theory wherein the ``live'' gauge field exchanges
can neither derail matter particles from their propagation paths nor
create virtual pairs from the vacuum. Any derailment occurs only on a
sudden-impulse basis and corresponds to the presence of cusps on the
propagation contour.

One may now ask whether such a procedure is sufficient to accomplish
the full factorization of the soft physics entering a given process,
and to what extent the resulting construction depends on particular
gauge choices. The answer to the first part of the question is implicit
in the methodology by which Sudakov-like suppression factors (Sudakov
logarithms) were successfully derived in Refs. \cite{GKKS97,KK98}.
As it turns out, the relevant, nonperturbative expression that factorizes
out corresponds to a LLA result which is swept by an appropriate
renormalization-group (RG) controlled running. Confronting the second
part of the question, on the other hand, has not been directly addressed
to before and will become one of the focal points in this investigation,
as we intend to elucidate the precise role played by the Feynman gauge
in our computations.

On the physical front, our present effort will direct itself towards the
derivation of the Sudakov-like suppression factor for the four-point
process associated with the wide angle ``elastic scattering'' involving
two {\it off-mass-shell} spin-1/2 matter particles in a non-abelian
gauge theory. Looking, at the same time, beyond this specific objective,
we view the present investigation as constituting an attempt to connect
parton-model based factorization schemes
\cite{DDT80,LB80,ER80,CS81,CZ84,ILS84,BS89,CSS89,LS92,BJKBS95,Szc96,SSK98}
with results founded in basic field theoretical calculations.
Our discussion will therefore naturally connect to previous works on
the understanding of Sudakov effects in elastic scattering
\cite{SS94,KOS98}, though the present approach is generically different
from those, as here open, i.e., Wilson lines of finite length are not
used in correspondence with the OPE but as fundamental ingredients of
the formalism.

{\bf 2.} We commence our exposition by displaying the basic formulas, in
worldline form, entering the analysis to follow.
Working in Euclidean space we introduce the four-point function
\begin{eqnarray}
\everymath{\displaystyle}
  \frac{\delta ^{4}}
       {\delta\overline{\eta}^{f}_{i}(x_{1}) \,
        \delta\overline{\eta}^{f}_{j}(x_{2}) \,
        \delta\eta ^{f}_{i'}(y_{1}) \,
        \delta\eta ^{f}_{j'}(y_{2})}
  \ln Z ( \overline{\eta}, \eta )|_{\overline{\eta}=\eta =0}
& = &
  {\cal M}\left(x_{1},x_{2};y_{1},y_{2}\right)^{ii'}_{jj'} \,
\nonumber \\
&&  - {\cal M}\left(x_{1},x_{2};y_{2},y_{1}\right)^{ij'}_{ji'} \; ,
\label{eq:4pf}
\end{eqnarray}
where $Z(\overline{\eta},\eta )$ is the partition function for the
non-abelian gauge field theory with spin-1/2 matter fields, whose
sources are $\overline{\eta},\,\eta$, whereas $f$ denotes flavor and
$i,j\ldots$ are group representation indices.

The expression for the fermion-fermion amplitude in the worldline
formalism is given by
\begin{eqnarray}
  {\cal M}^{ii'}_{jj'} \,
=\,
  \displaystyle{\sum_{C^I_{x_1,x_2}}} \,
  \displaystyle{\sum_{C^{II}_{x_2,y_2}}} \,
  I\left[\dot{x}^I\right] \, I\left[\dot{x}^{II}\right]
&&
  \Biggl\langle {\rm P} \exp
  \left[
        ig \int_{0}^{T_1} \, d\tau \, \dot{x}^I(\tau ) \cdot
        A\left(x^I(\tau)\right)
  \right]_{ii'}
\nonumber \\
&&
\times
  {\rm P} \exp \left[
          ig \int _{0}^{T_2} \, d\tau \, \dot{x}^{II}(\tau) \cdot
          A\left(x^{II}(\tau)\right)
               \right]_{jj'}\Biggr\rangle _A^{\rm conn}
\; ,
\label{eq:ferferampl}
\end{eqnarray}
where ${\langle\ldots\rangle}_{A}$ denotes functional averaging in the
gauge field sector with whatever this entails (ghosts, gauge choice
prescription, Dirac determinant, etc.) and, where
\begin{eqnarray}
  \displaystyle{\sum_{C^{I}_{x,y}}}I[\dot{x}] \,
\equiv
&&
  \int_{0}^{\infty} dT
  \displaystyle{\int_{\stackrel{x(0)=x}{x(T)=y}}} \,
  {\cal D}x(\tau )
  \int_{}^{}{\cal D}p(\tau ) \,
  {\rm P} \exp \left[ - \int_{0}^{T} \, d\tau
                     \left( i p(\tau ) \cdot \gamma \, + \, M \right)
               \right]
\nonumber \\
&& \times
  \exp \left[ i \int_{0}^{T} \, d\tau \,
  p(\tau ) \cdot \dot{x}(\tau )\right] \; .
\label{eq:parint}
\end{eqnarray}

\begin{figure}
\vspace{1.5 true cm}
\begin{picture}(0,40)
  \put(-65,-300){\psboxscaled{600}{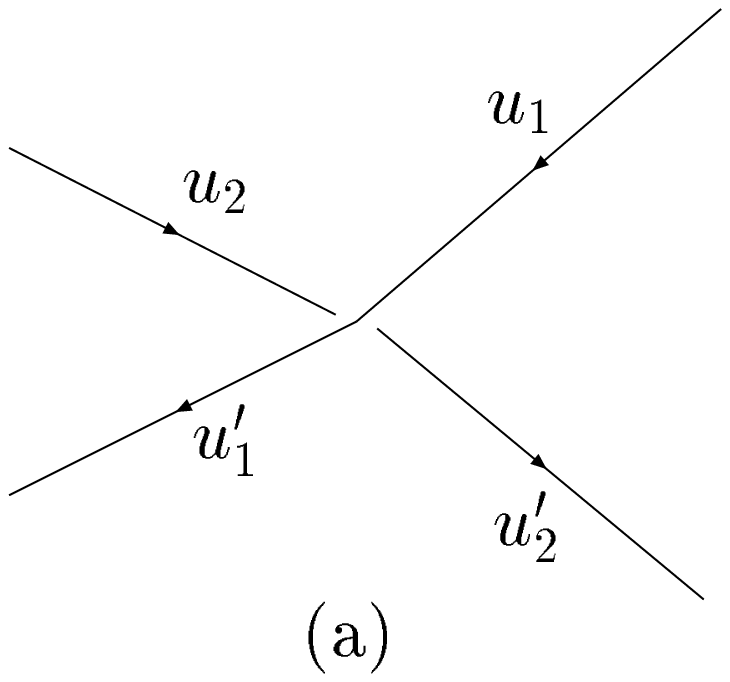}}
\end{picture}
\begin{picture}(0,40)
  \put(170,-300){\psboxscaled{600}{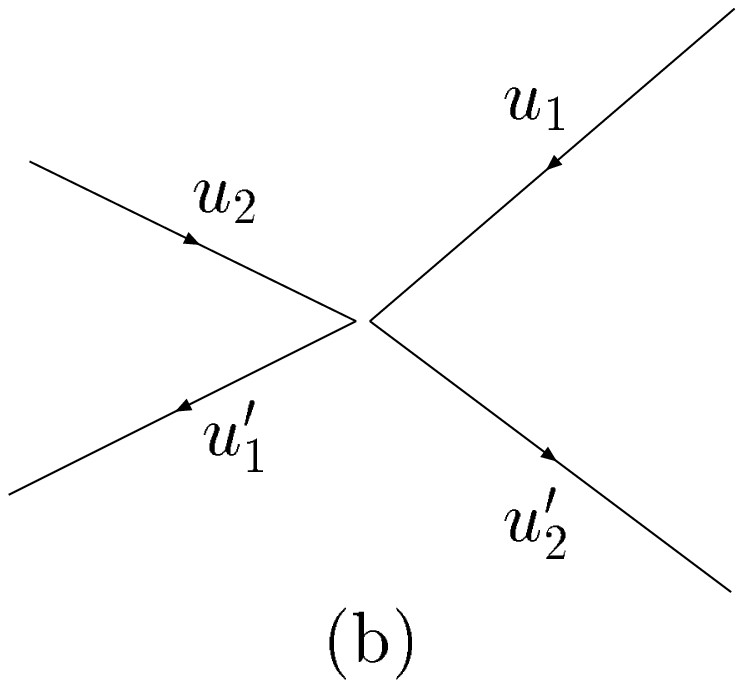}}
\end{picture}
\vspace{1.5 true cm}
\caption[fig:scatt]
{\tenrm  Worldline configurations for forward high-energy scattering
         of spin-$1/2$ matter particles with associated four-velocities.
         (a) shows the configuration which corresponds to particles
         being derailed via their mutual interaction on a sudden-impulse
         basis.
         (b) shows the double-cusped worldline configuration which mixes
         with the previous case under renormalization-group evolution.
\label{fig:scatter}}
\end{figure}

The configuration to be studied in this work pertains to two incoming
worldlines of {\it finite} extent with four-velocity vectors $u_{1}$
and $u_{2}$, respectively, which are derailed via a mutual interaction
operating on a sudden-impulse basis. The latter induces a cusp on each
line so that the colliding fermions exit with four-velocities
$u_{1}^{\prime}$ and $u_{2}^{\prime}$.
This situation is depicted in Fig. \ref{fig:scatter}a and corresponds,
when the matter particles are on-mass-shell, to an elastic-scattering
process at very high energies and large momentum transfers, i.e., to a
process that probes the limit
$s\rightarrow\infty$, $|t|\rightarrow\infty$ at fixed ratio $s/t$.
The on-mass-shell version has already been considered in the context of
the worldline formalism \cite{KK98} and has led to the derivation of the
on-mass-shell expression, giving rise to a Sudakov-like suppression
factor for the elastic amplitude.

%
\input psbox.tex
%

In the present investigation, the spin-1/2 entities entering the
four-point process are taken to be off-mass-shell. In our formalism the
off-mass-shellness is of the order of $1/\sigma$, where $\sigma$ stands
for the length of each (broken) straight-line path. Indeed, the finite
length of the matter particle's propagation contour serves to cut-off
all gauge field modes with momentum less than $1/\sigma$, participating
in its full, on-mass-shell description.

A useful kinematical parametrization for the on-mass-shell problem is
\begin{eqnarray}
&&
  p_{1}
=
  \left(\sqrt{Q^{2} + M^{2}},0,0,Q\right) ,
  \hspace{1cm}
  p_{2}
=
  \left(\sqrt{Q^{2} + M^{2}},0,0,-Q\right)
\nonumber \\
&&
  p'_{1}
=
  \left(\sqrt{Q^{2} + M^{2}},0,Q\sin\theta ,Q\cos\theta \right) ,
  \hspace{1cm}
  p'_{2}
=
  \left(\sqrt{Q^{2} + M^{2}},0,-Q\sin\theta,-Q\cos\theta \right) \; .
\label{eq:kinpar}
\end{eqnarray}
In turn, the variables $s$ and $t$ are given, respectively, by
\begin{equation}
  s \,
= \,
  \left(p_{1} + p_{2}\right)^{2} \,
= \,
  4\left(Q^{2} + M^{2}\right)
\end{equation}
and
\begin{equation}
  t \,
= \,
  \left(p_{1} - p'_{1}\right)^{2} \,
= \,
  - 2 Q^{2} \left(1 - \cos\theta \right) \; .
\end{equation}
The limit $s,|t|\rightarrow\infty$ with $s/t$ fixed is now taken in
the sense that $|Q|\rightarrow\infty$, and $\theta$ is hold fixed.
Though the above relations have only an {\it implicit} meaning for the
off-mass-shell situation in consideration here, we shall continue to
employ them in order to characterize the kinematical region under study
by a single large-momentum scale, namely $Q$. A precise characterization
of the (effective) mass parameter $M$, entering the off-mass-shell
analysis, will be made in the end. It should, in any case, be
anticipated that the off-mass-shellness will somehow be related to an
IR cutoff.

With the above definitions/clarifications in place, let us now introduce
the various scales that will enter our analysis. At the very top, we
place the c.m.-energy/momentum-transfer $|Q|$ at and above which the
truly hard, perturbative, physics acts. At the very bottom lies an IR
scale $\mu_{\rm IR}$ which marks the point from which our RG-running
commences. Note that, to the extent that we shall rely on perturbative
estimates of the various quantities entering the RG equation, we must
demand that $\mu_{\rm IR}\gg \Lambda _{\rm QCD}$. Finally, an in-between
scale $\Lambda$ serves to isolate that subsector of the full theory
which is associated with the (almost everywhere) straight-line
propagation of matter particles. We shall refer to the latter as the
{\it eikonal} (sub)sector. Clearly, all the potentially factorizable
soft physics reside in energy scales below $|Q|$ and in this sense
$\Lambda$ can ``run'' in the interval $[\mu _{\rm IR},\,|Q|]$. It is
worth noting that $|Q|$ also provides a measure for angles between a
pair of four-velocity vectors that enter and exit a given cusped
formation.

{\bf 3.} Generically speaking, the factorization of a quantity $W$ of
physical interest, associated with an (energy) separation scale
$\Lambda$, is based on a casting of the form
\begin{equation}
  W
\, = \,
  W_{\rm SOFT}W_{\rm HARD}\, + \, {\cal O}(1/\Lambda ) \; .
\label{eq:fact}
\end{equation}
In the framework of QCD and the language of Feynman diagrams, such an
assertion calls for intricate considerations
\cite{CS81,CSS89,BS89,SS94,KOS98,Li97}
which take into account contributions from homogeneously soft gluons,
on the one hand, and collinear gluons (in jets), on the other.

To the extent that a problem, such as the one in hand, involves a
large momentum scale $Q$ (representing, e.g., a large-momentum
transfer), it becomes convenient in order to extract the asymptotic
behavior of $W$ as $|Q|\rightarrow\infty$ to work with the quantity
$\frac{d}{d\ln Q^{2}}W$, whose independence from $\Lambda$, at
${\cal O}(1/\Lambda)$, leads to the RG equation
\begin{equation}
   \left(\mu\frac{\partial}{\partial\mu}
  + \beta\frac{\partial}{\partial g}
    \right)
    \frac{d}{d\ln Q^{2}}\ln W_{\rm HARD}
=
  - \left(\mu\frac{\partial}{\partial\mu}
  + \beta\frac{\partial}{\partial g}
    \right)
    \frac{d}{d\ln Q^{2}}\ln W_{\rm SOFT}
=
    \Gamma _{\rm S} \; ,
\label{eq:RGE}
\end{equation}
where $\Gamma _{\rm S}$ stands for anomalous dimensions pertaining to
$\frac{d}{d\ln Q^{2}}W_{\rm SOFT}$,
and where we have replaced $\Lambda$ by $\mu$ adhering to the
conventional choice for the running scale.

Our obvious intention is to calculate anomalous dimensions associated
with the configuration of Fig. \ref{fig:scatter}a, wherein what is
emitted or absorbed by the straight-line segments are uniformly soft
gluons, with respect to $\Lambda$, i.e., employing a no-impulse
approximation.
The existence of these anomalous dimensions can be understood on
the basis that, from the perspective of the IR cut-off $\mu _{\rm IR}$,
the upper momentum scale $\Lambda$ appears as being infinitely remote --
after all, it is the ratio between UV and IR cutoffs that really
matters, as far as RG-running is concerned. To what extent, on the other
hand, the physics operating within this eikonal subsector can fully
account for the long-distance behavior of $W$ constitutes an open
question which will not be addressed here.

In what follows, we shall focus our efforts on the disconnected
four-point function which we denote by $(W_{1})^{ii'}_{jj'}$.
Non-abelian group considerations lead us to define the following
invariant quantities, see, e.g., \cite{Li88,Kor94},
\begin{equation}
  W_{1}^{(a)}
\equiv
  {\langle{\rm tr} P_{I} \, {\rm tr} P_{II}\rangle}_{A}
\, = \,
  \delta _{ii'}\delta _{jj'} \,
\left(W_{1}\right)^{ii'}_{jj'}
\label{eq:inv1}
\end{equation}
and
\begin{equation}
  W_{1}^{(b)}
\equiv
  {\langle{\rm tr}(P_{I}P_{II})\rangle}_{A}
\, = \,
  \delta _{ij'}\delta _{ji'}
\left(W_{1}\right)^{ii'}_{jj'} \; ,
\label{eq:inv2}
\end{equation}
where $P_{I}$ denotes a line configuration parametrized by
$\dot{x}^{I}(s)$ and $P_{II}$ another one parametrized by
$\dot{x}^{II}(s)$.
In the situation shown in Fig. \ref{fig:scatter}a, of course,
$\dot{x}^{I}(s) = u_{1}$ in $[0,s_{1}]$, $\dot{x}^{I}(s) = u'_{1}$ in
$\left[s_{1},\sigma\right]$, $\dot{x}^{II}(s) = u_{2}$ in
$\left[0,s_{2}\right]$ and
$\dot{x}^{II}(s) = u'_{2}$ in $\left[s_{2},\sigma\right]$, where
$s_{1}$ and $s_{2}$ mark the derailment points on the respective
contours.

It follows that
\begin{equation}
  \left(W_{1}\right)^{ii'}_{jj'}
\, = \,
    \frac{N\, W_{1}^{(a)} - W_{1}^{(b)}}{\displaystyle N(N^{2}-1)} \,
    \delta _{ii'}\delta _{jj'}
  + \frac{N\, W_{1}^{(b)}- W_{1}^{(a)}}{\displaystyle N(N^{2}-1)}
    \delta _{ij'}\delta _{ji'} \; .
\label{eq:kinbran}
\end{equation}

Referring to the invariant quantities $W_{1}^{(a,b)}$, the
RG equation (\ref{eq:RGE}) reads, up to possible mixing with other
operators,
{\everymath{\displaystyle}
\begin{eqnarray}
&&
    \left(\mu\frac{\partial}{\partial\mu}
  + \beta\frac{\partial}{\partial g}
    \right)
    \frac{d}{d\ln Q^{2}}
    \ln \left(W_{1}^{(a,b)}\right)_{\rm HARD}
\nonumber \\
&& \quad =
  - \left(\mu\frac{\partial}{\partial\mu}
  + \beta\frac{\partial}{\partial g}
    \right)
    \frac{d}{d\ln Q^{2}}\ln\left(W_{1}^{(a,b)}\right)_{\rm SOFT}
=
  \Gamma _{\rm S}^{(a,b)} \; .
\label{eq:RGEW}
\end{eqnarray}}
But, provided we find the appropriate anomalous dimensions associated
with the soft factor, then $\frac{d}{d\ln Q^{2}}W_{1}^{(a,b)}$ can be
determined via a two-step procedure which addresses itself first to its
soft and second to its hard component.

The crucial point now is that the worldline scheme allows us to
compute unambiguously anomalous dimensions associated with the
homogeneously soft gluon region. In our case, the relevant configuration
is that represented graphically in Fig. \ref{fig:scatter}a. As it will
turn out, these will be the anomalous dimensions denoted
$\Gamma _{\rm S}$ in (\ref{eq:RGEW}), or stated equivalently: the soft
physics that we shall be in the position to factorize pertains to the
eikonal subsector. Our first task, however, is to establish that it is
enough to perform the relevant calculation in the Feynman gauge.

{\bf 4.} Consider the perturbative expansion of the expectation value,
with respect to the gauge field sector, entering the expression for the
four-point function under study, cf. Eq.~(\ref{eq:ferferampl}),
{\it albeit} now for the disconnected part. A typical term, which
involves an interaction point, is of the form
\begin{equation}
  [V]_{jj'}^{ii'}
=
  (ig)^{2} \int_{0}^{T_1} d\tau
  \int_{0}^{T_2} d\tau '
  \dot{x}^{I}_{\mu}(\tau ) \dot{x}^{II}_{\nu}(\tau ')
  {\left\langle A_{\mu}\left(x^{I}(\tau )\right)_{ii'}
                A_{\nu}\left(x^{II}(\tau )\right)_{jj'}
  \right\rangle}_A \; .
\label{eq:typterm}
\end{equation}

Let us first treat the case of covariant gauge choices. Anticipating the
fact that the eikonal subsector of the full theory has its own UV
domain, we work with a (dimensionally) regularized form of the
propagator, which, generically, reads
\begin{equation}
  D_{\mu\nu}(|x|)
\, = \,
  \mu ^{4-D} \int {}^{} \frac{d^{D}k}{(2\pi )^{D}} \,
  {\rm e}^{-i k \cdot x}
  \frac{1}{k^{2}}\left[\delta _{\mu\nu} - (1-\xi )
                     \frac{k_{\mu} k_{\nu}}{k^{2}}
                 \right]
=
  D^{({\rm F})}_{\mu\nu}\, - \, (1-\xi ) D^{(\xi)}_{\mu\nu} \; ,
\label{eq:Feybopro}
\end{equation}
where $D^{({\rm F})}_{\mu\nu}$ stands for the Feynman gauge propagator.

We readily determine
\begin{equation}
  D^{({\rm F})}_{\mu\nu}(|x|)
\, = \,
  \delta _{\mu\nu}
  \frac{\mu ^{4-D}}{4\pi ^{D/2}} \,
  \Gamma \left(\frac{D}{2} - 1 \right)\frac{1}{|x|^{D-2}}
\label{eq:bopropFey}
\end{equation}
and
\begin{equation}
  D^{(\xi)}_{\mu\nu}(|x|)
\, = \,
  \frac{\mu ^{4-D}}{4\pi ^{D/2}} \,
  \Gamma \left(\frac{D}{2} - 1 \right)
  \frac{1}{2}
  \left[
          \frac{\delta _{\mu\nu}}{|x|^{D-2}}
        - (D-2) \frac{x_\mu x_\nu}{|x|^D}
  \right] \; .
\label{eq:boproxi}
\end{equation}
Consequently, we have
\begin{equation}
  D_{\mu\nu}(x-x')
\, = \,
  \frac{\mu ^{4-D}}{4\pi ^{D/2}} \,
  \Gamma \left(\frac{D}{2} - 1 \right)
  \left[ \frac{\delta _{\mu\nu}}{|x-x'|^{D-2}}
        -\frac{1}{2}(1 - \xi)
         \frac{1}{D-4}
         \frac{\partial}{\partial x_\mu}
         \frac{\partial}{\partial x'_\nu}
         \frac{1}{|x-x'|^{D-4}}
  \right] \; .
\label{eq:fullbopro}
\end{equation}

Inserting in Eq. (\ref{eq:typterm}) the second term in the square
brackets of this expression, leads to integrals having the typical form
\begin{eqnarray}
&&
  \frac{\mu ^{4-D}}{D-4}\int_{x_1}^{y_1}dx_{\mu}
                        \int_{x_2}^{y_2}dx'_{\nu}
  \frac{\partial}{\partial x_\mu}
  \frac{\partial}{\partial x'_\nu}|x-x'|^{4-D}
\nonumber \\
&& \quad
=  \frac{\mu ^{4-D}}{D-4}
   \left[
          |y_{1}-y_{2}|^{4-D} - |x_{1}-y_{2}|^{4-D}
         -|y_{1}-x_{2}|^{4-D} - |x_{1}-x_{2}|^{4-D}
   \right] \; ,
\label{eq:typint}
\end{eqnarray}
where $x_{1},\,x_{2}$ denote the initial and $y_{1},\,y_{2}$ the final
points of the two fermionic paths. The latter are obliged to pass
through the respective deflection (equivalently, interaction) points
$z_{1}$, $z_{2}$ under the condition that they respect the kinematics of
the process.

Asymptotically, the length of each of the four branches (two per cusped
contour) is very large, the order of which will be denoted by $|L|$.
Moreover, the projection of a given branch on another in the vicinity of
the deflection point is of order $Q^{2}/m^{2}$. As a consequence, the
rhs of the above relation becomes
\begin{equation}
  \frac{1}{D-4}\left[\mu |L|\frac{|Q|}{m}\right]^{4-D}
\longrightarrow
  \ln(\mu |L|) \, + \, \ln\frac{|Q|}{m}
\label{eq:rhs18}
\end{equation}
and hence does not yield an anomalous-dimension\footnote{Clearly,
$\ln(\mu |L|)$ furnishes the divergent logarithmic factor in the
eikonal subsector.} contribution to
$\frac{d}{d\ln Q^{2}}[V]^{ii'}_{jj'}$.
One surmises that, for an arbitrary covariant gauge choice, the
anomalous dimensions in the eikonal subsector are exclusively
associated with the expression furnished by the Feynman propagator.

Turning our attention now to an axial gauge, we write
\begin{equation}
  D_{\mu\nu}(|x|)
\, = \,
  D_{\mu\nu}^{(F)}(|x|) - D_{\mu\nu}^{(\eta )}(|x|) \; ,
\label{eq:axial1}
\end{equation}
where
\begin{equation}
  D_{\mu\nu}^{(\eta )}(|x|)
\, = \,
  \mu ^{4-D} \int_{}^{}\frac{d^{D}k}{(2\pi )^{D}} \,
  {\rm e}^{-i k \cdot x}
  \left[
          \frac{\eta _{\mu}k_{\nu} + \eta _{\nu}k_{\mu}}{\eta\cdot k}
        - \eta ^{2}\frac{k_{\mu}k_{\nu}}{(\eta\cdot k)^{2}}
  \right] \; .
\label{eq:axial2}
\end{equation}

The last term in the square brackets amounts to a double derivative
action and leads to the same result as the $\xi$-dependent part of the
covariant gauge propagator. The remainder, when inserted into Eq.
(\ref{eq:typterm}), leads to the following expression
\begin{eqnarray}
&&
  \int_{0}^{\infty} ds
  \int_{x_{1}}^{y_{1}} dx_{\mu} \int_{x_{2}}^{y_{2}} dx'_{\nu}
  \left[
          \eta _{\mu}\frac{\partial}{\partial x'_{\nu}}
        + \eta _{\nu}\frac{\partial}{\partial x'_\mu}
  \right]
         |x-x'+s\eta |^{D-2}
\nonumber \\
&& \quad =
  2 \int_{x_{1}}^{y_{1}} dx_{\mu}
  \left[\int_{-\infty}^{y_{2}} dw_{\mu}
        \frac{1}{|x-w|^{D-2}} \, - \,
        \int_{-\infty}^{x_{2}} d\bar{w}_{\mu}
        \frac{1}{|x-\bar{w}|^{D-2}}
  \right] \; ,
\label{eq:remainder}
\end{eqnarray}
where
$
 w_{\mu}
\equiv
 y_{2\mu} - s\eta _{\mu}$, $\bar{w}_{\mu}
\equiv
 x_{2\mu} - s\eta _{\mu}
$.

Now, the main contribution in the above two integrals comes,
respectively, from the points
$
 x_{\mu}
\simeq
 w_{\mu}
$,
$
 x_{\mu}
\simeq
 \bar{w}_{\mu}
$
and is of order $\ln (\Lambda )\ln (Q)$. However, the corresponding
terms register in the overall expression with opposite signs and
consequently cancel each other.

Generalizing to the full disconnected four-point function
$[W_{1}]^{ii'}_{jj'}$, one concludes that, for an arbitrary choice of
gauge condition, the leading behavior giving rise to anomalous
dimensions for $\frac{d}{d\ln Q^{2}}[W_{1}]^{ii'}_{jj'}$ comes from the
insertion of the Feynman propagator in the full expression for the
four-point function. It is, moreover, of crucial importance to observe
that the relevant singularity structure associated with interaction
points {\it will be invariably picked up by the (broken) straight-line
configurations}. Indeed, as becomes evident from
Eq. (\ref{eq:bopropFey}), the main contribution to the terms involving
two different branches exiting a point of interaction comes from the
immediate vicinity of the latter\footnote{It is perhaps helpful to
rewrite the relevant factor in Eq. (\ref{eq:bopropFey}) as
$\displaystyle\frac{1}{|x(\tau )-x(\tau ')|^{D-2}}$
and let $\tau$, $\tau '$ run, respectively, along branches with
different four-velocities.}. In addition, such singularities will be
picked up by an arbitrary path, entering the full expression for
$[W_{1}]^{ii'}_{jj'}$ on account of the constraint imposed by the
momentum transfer injected at the cusp. This implies that the eikonal
sector, we have isolated via the worldline casting of the system,
automatically factorizes carrying with it the relevant, overall
renormalization factor. For four-point processes, in general, such
factors, induced by singularities associated with interaction points,
have been identified as being of the cusp \cite{KR86} or the cross
\cite{Kor94} type. For the present case, in which a large momentum
transfer is involved, our concern will be with cusp-type singularities.

{\bf 5.} The preceding analysis has led us to the specific task of
determining anomalous dimensions for the quantity
$\frac{d}{d\ln Q^{2}}[(W_{1})_{\rm EIK}]^{ii'}_{jj'}$.
The explicit expression for $[(W_1)_{\rm EIK}]^{ii'}_{jj'}$ to
${\cal O}(g^{2})$ reads
\begin{eqnarray}
  \left[(W_{1})_{\rm EIK}\right]^{ii'}_{jj'}
=
   \delta _{ii'}\delta _{jj'}
& + &
  (ig)^{2}\int_{0}^{\sigma} d\tau
          \int_{0}^{\sigma} d\tau '
  \Biggl\{
          C_{\rm F}\delta _{ii'}\delta _{jj'}
  \Bigl[
          u_{1\mu}u_{1\nu}
          D_{\mu\nu}(|\tau u_{1} - \tau 'u_{1}|)
    + (1\leftrightarrow 2)
\nonumber \\
  & + & \ldots
  \Bigr]
    + t^{a}_{ii'}t^{a}_{jj'}
  \Bigl[
          u_{1\mu} u_{2\nu}
          D_{\mu\nu}(|\tau u_{1} - \tau ' u_{2}|)
    + \ldots
  \Bigr]
\Biggr\}
\label{eq:eikonal}
\end{eqnarray}
where the eclipse denotes terms which comprise combinations of each
worldline branch carrying four-velocity
$u_{i}^{(\prime )}=p_{i}^{(\prime )}/M$
either with itself (second bracket) or with the other three branches,
while involving, at the same time, a gluon exchange across the point of
interaction.
Note that for those terms which contain $u_{i}u_{j}^{\prime}$, the
argument of $D_{\mu\nu}$ has the reverse sign.
Note also that, as we have already established, the gluon propagator
can be taken in the Feynman gauge without loss of generality.

Our basic calculational task amounts to dealing with integrals of the
form
\begin{equation}
  I^{(\pm)}
\equiv
  u_{i} \cdot u_{j} \int_{0}^{\sigma} d\tau
  \int_{0}^{\sigma} d\tau' D(\tau u_{i} \pm \tau' u_{j})
\label{eq:integral1}
\end{equation}
where $u_{i,j}$ stands for $u_{1}^{(\prime )}$ and  $u_{2}^{(\prime )}$.

One finds, as $D\rightarrow 4$,
\begin{equation}
  \int_{-\sigma}^{+\sigma} d\tau
  \int_{-\sigma}^{+\sigma} d\tau ' \,
  u_{i} \cdot u_{j} \, D(|u_{i} \tau \pm u_{j} \tau '|)
\, = \,
  \frac{1}{4\pi ^{2}} \,
  {\left(
         \frac{\mu ^2}{\tilde{\lambda}^{2}}\pi
  \right)}^{\epsilon} \,
  \frac{1}{2\epsilon} \,
  f^{(\pm)}_{4 - 2\epsilon}(w_{ij}) \; ,
\label{eq:propd4}
\end{equation}
where we have used
$
 w_{ij}
\equiv
 u_{i}\cdot u_{j}
$,
$
 \tilde{\lambda}
\equiv
 1/\sigma
$
and
$\epsilon=4-D(>0)$.
Here
\begin{equation}
  f^{(+)}_{4-2\epsilon}
=
       2w^{2} F\left(1,1;\frac{3}{2};1-w^{2}\right) \,
  - \, [2(1+w)]^{\epsilon} wF\left(1,1;\frac{3}{2};\frac{1-w}{2}\right)
\label{eq:fplus}
\end{equation}
and
\begin{equation}
  f^{(-)}_{4-2\epsilon}
=
      2w^{2} F\left( 1,1;\frac{3}{2};w^{2} \right) \,
  +\, [2(1-w)]^{\epsilon} wF\left( 1,1;\frac{3}{2};\frac{1-w}{2} \right)
\; ,
\label{eq:fminus}
\end{equation}
with $w=u_{1} \cdot u_{2}/|u_{1}||u_{2}|$.

Taking now into account that, in the fundamental representation,
\begin{equation}
  t^{\alpha}_{ii'}t^{\alpha}_{jj'}
=
 - \frac{1}{2N}\delta _{ii'}\delta _{jj'}
 + \frac{1}{2} \delta _{ij'}\delta _{ji'} \; ,
\label{eq:gener}
\end{equation}
we determine
\begin{equation}
  (W_{1}^{(a)})_{\rm EIK}
\, = \,
  N^{2}C_{11} \, + \, NC_{12}
\label{eq:W_1^a}
\end{equation}
and
\begin{equation}
  (W_{1}^{(b)})_{\rm EIK}
\, = \,
  NC_{11} \, + \, N^{2}C_{12}
\label{eq:W_1^b}
\end{equation}
with $C_{11}$ and $C_{12}$ remaining to be calculated. In fact, we are
actually interested in
$\frac{\partial}{\partial \ln Q^{2}}C_{1i}$
and, as far as the leading-order estimate is concerned, in the singular
terms entering the respective (asymptotic) expressions from which the
anomalous dimensions will be extracted. For that reason, we shall
restrict ourselves to a minimal exposition of relevant mathematical
manipulations by explicitly treating only the case of $C_{11}$.

Setting $w_{ij}=\cos\phi _{ij}$ and going over to Minkowski space
($\phi _{ij}\rightarrow -i\gamma _{ij}$), we obtain (by using Eqs.
(\ref{eq:propd4})-(\ref{eq:fminus}), removing pole terms via the
$\overline{\mbox{MS}}$ subtraction scheme, and going to the limit
$\epsilon\rightarrow 0$)
\begin{eqnarray}
  C_{11}
& = &
  1 - \frac{\alpha _{\rm s}}{\pi}
\Biggl\{
        \ln (\mu\sigma |u|)
        \left[
                2C_{\rm F}(\gamma _{11'} {\rm coth}\gamma _{11'} - 2)
              - \frac{1}{N}
            \left(
                    (i\pi -\gamma _{12}){\rm coth}\gamma _{12}
                  + \gamma _{12'}{\rm coth}\gamma _{12'}
            \right)
        \right]
\nonumber \\
&& \quad +
           C_{\rm F}h^{(+)}(\gamma _{11'})
         - \frac{1}{2N}
           \left[
                 h^{(+)}(\gamma _{12'}) + h^{(-)}(\gamma _{12}
           \right]
         - 4 C_{\rm F}
\Biggr\}
         + {\cal O}(g^4) \; ,
\label{eq:C_11}
\end{eqnarray}
where
\begin{equation}
  h^{(\pm )}(w)
\simeq
   \pm \frac{1}{4} \ln ^{2}\left(\pm 2w\right) \,
  + \, {\cal O}\left(\frac{1}{w}\right) \; .
\label{eq:fun_h}
\end{equation}
These functions enter, typically, off-mass-shell expressions in the
worldline scheme \cite{KK98}.

In the asymptotic regime that is of interest to us, we get from
Eq. (\ref{eq:kinpar}) and subsequent transcription to Minkowski space,
\begin{equation}
  \gamma _{12}
\simeq
  \ln\left(\frac{2Q^{2}}{M^{2}}\right) , \;
  \gamma _{11'}
\simeq
   \ln\left(\frac{2Q^{2}}{M^{2}}\right)
  +\ln\left(\sin ^{2}\frac{\theta}{2}\right) , \;
  \gamma _{12'}
\simeq
   \ln\left(\frac{2Q^{2}}{M^{2}}\right)
  +\ln\left(\cos ^{2}\frac{\theta}{2}\right) \; .
\label{eq:gammas}
\end{equation}

We, thereby, obtain
\begin{equation}
    \frac{\partial}{\partial \ln Q^{2}} \, C_{11}
\simeq
  - \frac{\alpha _{\rm s}}{\pi}
\left[
        2C_{\rm F} \ln \left(\mu\bar{\sigma}\right)
      +  C_{\rm F} \ln \left(2\sin ^{2}\frac{\theta}{2}\right)
      - \frac{1}{N}\ln \left(2\cos ^{2}\frac{\theta}{2}\right)
      - \frac{i\pi}{2N}
\right] \; ,
\label{eq:evolC_11}
\end{equation}
where
$
 \bar{\sigma}
\equiv
 |u|\sigma\left(\frac{2Q^{2}}{M^{2}}\right)^{1/2}
$
properly defines the lowest scale of the RG running, thereby
facilitating the identification $\bar{\sigma}\equiv\mu _{\rm IR}^{-1}$.

From the above result, we read off the relevant contribution to the
anomalous dimensions, namely the coefficient in front of
$\ln (\mu\bar{\sigma})$: $-2C_{\rm F}\frac{\alpha _{\rm s}}{\pi}$. On
the other hand, it turns out that there is no contribution to the
anomalous dimensions from $\frac{\partial}{\partial\ln Q^{2}}C_{12}$.

{\bf 6.} The reason for carrying all along the index $1$ on $W$ is that
under the RG running the four-point function associated with the
scattering process mixes with
$[(W_{2})_{\rm EIK}]^{ii'}_{jj'}$ which corresponds to the Wilson-line
arrangement shown in Fig. \ref{fig:scatter}b
\cite{Bra81}. One may represent this quantity by
\begin{eqnarray}
  \left[(W_{2})_{\rm EIK}\right]^{ii'}_{jj'}
& = &
  \Biggl\langle {\rm P}
  \exp\left[
              ig\int_{-\sigma}^{0} d\tau  u_{1}  \cdot A(\tau u_{1})
            + ig\int_{0}^{\sigma}  d\tau  u'_{2} \cdot A(\tau u'_{2})
      \right]_{ij'}
\nonumber \\
&& \quad \times
      {\rm P} \exp
      \left[
              ig\int_{-\sigma}^{0} d\tau u_{2}  \cdot A(\tau u_{2})
            + ig\int_{0}^{\sigma}  d\tau u'_{1} \cdot A(\tau u'_{1})
      \right]_{ji'}\Biggr\rangle _{A} \; ,
\label{eq:W_2}
\end{eqnarray}
with the agreement that it acquires literal meaning in the Feynman
gauge.

Corresponding invariant functions $(W_{2}^{(a,b)})_{\rm EIK}$ can now
be introduced in a similar fashion as in Eqs. (\ref{eq:inv1}),
(\ref{eq:inv2}). Moreover, we may write
\begin{equation}
  (W_{2}^{(a)})_{\rm EIK}\, = \, N^{2}C_{21} \, + \, N C_{22},\;\;\;
  (W_{2}^{(b)})_{\rm EIK}\, = \, N C_{21} \, + \, N^{2} C_{22}
\label{eq:C_2i}
\end{equation}
and face, for the determination of the $C_{2i}$, similar tasks to those
involved for the determination of $C_{1i}$. The encountered integrals
are the same as before. The end result is that we have a contribution to
the anomalous dimensions from
$\frac{\partial}{\partial \ln Q^{2}}C_{22}$
being equal to
$-2C_{\rm F}\frac{\alpha _{\rm s}}{\pi}$,
i.e., it is the same as for
$\frac{\partial}{\partial \ln Q^{2}}C_{11}$,
whereas
$\frac{\partial}{\partial \ln Q^{2}}C_{21}$
provides a null contribution.

Before carrying through the RG analysis, we need to adjust our formalism
in such a way as that it applies to the invariant quantities $W^{(i)}$,
($i=a,b$). To this end, let us set
\begin{equation}
  \tilde{W}_{\rm EIK}^{(i)}
\equiv
  \left(\begin{array}{c}W_{1}^{(i)}\\ W_{2}^{(i)}
\end{array}\right)_{\rm EIK}.
\label{eq:vecW_EIK}
\end{equation}
We define
\begin{equation}
  \tilde{F}_{\rm EIK}^{(i)} \left( \frac{\mu}{\mu_{\rm IR}},
                                   \frac{Q^{2}}{M^{2}}
                            \right)
\equiv
  \frac{d}{d\ln Q^{2}}
  \ln\tilde{W}_{\rm EIK}^{(i)}\left( \frac{\mu}{\mu_{\rm IR}},
                                     \frac{Q^{2}}{M^{2}}
                              \right)
=
  - \ln\left(\frac{\mu}{\mu_{\rm IR}}\right)\hat{\Gamma}^{(i)}_{\rm EIK}
  - \hat{D}^{(i)}_{\rm EIK},\;\;i=a,b,
\label{eq:F_EIK}
\end{equation}
where the quantities $\hat{\Gamma}^{(i)}_{\rm EIK}$ and
$\hat{D}^{(i)}_{\rm EIK}$
can be surmised from the expressions for the $C_{ij}$.

The RG equation, then, reads
\begin{equation}
  \left(
        \mu\frac{\partial}{\partial\mu} + \beta (g)
                                          \frac{\partial}{\partial g}
  \right)\tilde{F}_{\rm EIK}^{(i)}
\, = \,
   - 2 C_{\rm F} \,
     \frac{\alpha _{\rm s}}{\pi}\hat{\Gamma}^{(i)}_{\rm EIK} \; .
\label{eq:RGE/F_EIK}
\end{equation}
We readily determine that the associated initial conditions are given
by
\begin{equation}
  \tilde{F}_{\rm EIK}^{(i)}\left(
                                 \frac{\mu}{\mu_{\rm IR}},
                                 \frac{Q^{2}}{M^{2}}
                           \right)_{\mu =\mu_{\rm IR}}
=
  - \tilde{D}_{\rm EIK}^{(i)}(\alpha _{\rm s}(\mu _{\rm IR})) \; .
\label{eq:incoF_EIK}
\end{equation}
As we are primarily interested in isolating the leading-order result,
we write
\begin{equation}
\hat{F}^{(i)}_{\rm EIK}\left(\frac{\mu}{\mu _{\rm IR}},
                             \frac{Q^{2}}{m^{2}}
                       \right)
=
  - \frac{2C_{\rm F}}{\pi}\int_{\mu_{\rm IR}}^{\mu}
    \frac{d\tau}{\tau}\alpha _{\rm s}(\tau ) \, + \,
\mbox{nonleading terms}
\label{eq:LOF_EIK}
\end{equation}
in which the nonleading terms correspond to
$
   \tilde{D}_{\rm EIK}^{(a)}(\alpha _{\rm s}(\mu _{\rm IR}))
 + {\cal O}\left(\alpha _{\rm s}^{2}\right)
$.

According to our procedure, what we have factorized with respect to the
energy scale $\mu\in[\mu _{\rm IR},|Q|]$, is eikonal vs. non-eikonal
physics operative in this range. The latter contains configurations that
go beyond the no-impulse approximation. Referring to
Eq. (\ref{eq:RGEW}), after adjusting to the present situation, we have
\begin{equation}
    \frac{d}{d\ln Q^{2}}\ln (W_{1})^{(i)}_{\rm NON-EIK}
=
  - \frac{2C_{\rm F}}{\pi}\int_{\mu}^{|Q|}
    \frac{d\tau}{\tau}\alpha _{\rm s}(\tau )
\, + \, \mbox{nonleading terms},
\label{eq:W_nEIK}
\end{equation}
where now the nonleading terms are associated with initial conditions
pertaining to the running of the non-eikonal contributions with respect
to $\mu$.

Putting everything together, we finally obtain
\begin{equation}
    \frac{d}{d\ln Q^{2}}\ln(W_{1})^{(i)}
=
  - \frac{2C_{\rm F}}{\pi}\int_{\mu_{\rm IR}}^{|Q|}
    \frac{d\tau}{\tau}\alpha _{\rm s}(\tau )
\, + \, \mbox{nonleading terms} \; .
\label{eq:RGEfinW}
\end{equation}

Next, introducing the mass scale
$
 \bar{M}
\equiv
 \left(\mu _{\rm IR}^{2}Q^{2}\right)^{1/4}
=
 \left(\frac{M^{2}}{2\sigma ^{2}|u|^{2}}\right)^{1/4}
$,
which should be greater than $\Lambda _{\rm QCD}$ for the
perturbative analysis to be valid, we may express the leading order RG
result as follows
\begin{equation}
  \ln W_{1}^{(i)}
=
  - \frac{4 C_{\rm F}}{\beta _{0}}
  \left[
          \ln\frac{Q^{2}}{\Lambda _{\rm QCD}^{2}}
          \ln\ln\frac{Q^{2}}{\Lambda _{\rm QCD}^{2}}
        + \ln\frac{\bar{M}^{4}}{\Lambda_{\rm QCD}^{2}Q^{2}}
          \ln\ln\frac{\bar{M}^{4}}{\Lambda _{\rm QCD}^{2}Q^{2}} \; ,
  \right]
\label{eq:lnW1}
\end{equation}
which, after a number of routine algebraic manipulations, can be cast
into a form that helps isolating its dominant term. The relevant
expression then reads
\begin{eqnarray}
\everymath{\displaystyle}
      \ln W_{1}^{(i)}
&& =
    - \frac{8 C_{\rm F}}{\beta _{0}}
      \ln\frac{Q^{2}}{\Lambda _{\rm QCD}^{2}}
      \ln\ln\frac{Q^{2}}{\Lambda _{\rm QCD}^{2}}
    + \frac{8 C_{\rm F}}{\beta _{0}} \ln\frac{Q^{2}}{\bar{M}^{2}}
      \ln\ln\frac{Q^{2}}{\Lambda _{\rm QCD}^{2}}
\nonumber \\
&&
    - \frac{4 C_{\rm F}}{\beta _{0}}
      \ln\frac{\bar{M}^{4}}{\Lambda _{\rm QCD}^{2}Q^{2}}
      \ln\ln \left[
                   1 - 2\frac{\ln\frac{Q^{2}}{\bar{M}^{2}}}
                             {\ln\frac{Q^{2}}{\Lambda _{\rm QCD}^{2}}}
             \right] \; .
\label{eq:lnW_1fin}
\end{eqnarray}
Given the hierarchy of scales
$
 Q^{2}
>
 \bar{M}^{2}
>
 \frac{\bar{M}^{4}}{Q^2}
>
 \Lambda _{\rm QCD}^{2}
$,
we identify the first term on the rhs of this equation as the dominant
one. Further, it follows from the relation
$\sigma \sim \frac{m}{p^{2}-m^{2}}$,
which can be deduced from the original path integral, that the
effective mass scale entering our final results is related to the
off-mass-shellness: $\bar{M^{2}} \sim \left(p^{2}-m^{2}\right)$.

Just as in the case of the three-point function, we have calculated
earlier \cite{GKKS97}, one witnesses the fact that the off-mass-shell
result for the Sudakov suppression has an exponent twice as large as
the one for the on-mass-shell case \cite{Sen81}.
Moreover, it becomes obvious that the asymptotic, long-range behavior
of $W_{1}^{(a)}$ and $W_{1}^{(b)}$ is the same, so that in fact it
characterizes the long-range behavior of the total expression
$[W_{1}]^{ii'}_{jj'}$.
We have, in other words, extracted the Sudakov-like suppression factor
{\it for the four-point function as a whole}. This extends the
{\it hitherto} known result which refers to three-point vertices,
i.e., the one pertaining to form factors\footnote{Actually, the term
``Sudakov form factor'' is a misnomer since it does not account for
structure but merely for long-distance behavior of the vertex function
in the limit of large-momentum transfer.}. If, now, we wish to go to the
{\it connected} four-point function, then we must subtract a term
comprising the product of two Sudakov form factors. But, then, we shall
obtain precisely the same leading factor as in Eq. (\ref{eq:lnW_1fin}),
see, e.g., \cite{KK98,GKKS97}.

In conclusion, we have established the Sudakov behavior for four-point
processes through a computational process that has taken place within
the framework of a pure, field theoretical treatment with a
relatively painless effort.

\bigskip

\acknowledgements
The research of C. N. K. and S. M. H. W. was supported in part by
the EC Programme ``Training and Mobility of Researchers'' under
contract ERB FMRX-CT96-0008.

\newpage     


\begin{thebibliography}{999}
\bibitem{DDT80} Yu.L. Dokshitzer, D.I. Dyakonov, and S.I. Troyan,
                Phys. Rep. 58 (1980) 269.

\bibitem{Col89} J.C. Collins, in
                Perturbative Quantum Chromodynamics,
                ed. A.H. Mueller
                (World Scientific, Singapore, 1989),
                p. 573.

\bibitem{Mue81} A.H. Mueller,
                Phys. Rep. 73 (1981) 237; and references therein.

\bibitem{KK98} A.I. Karanikas and C.N. Ktorides,
               Phys. Rev. D 59 (1998) 016003.

\bibitem{KK92} A.I. Karanikas and C.N. Ktorides,
               Phys. Lett. B 275 (1992) 403;
               Phys. Rev. D 52 (1995) 5883.

\bibitem{KKS95} A.I. Karanikas, C.N. Ktorides, and N.G. Stefanis,
                Phys. Rev. D 52 (1995) 5898.

\bibitem{GKKS97} G. Gellas, A.I. Karanikas, C.N. Ktorides, and
                 N.G. Stefanis,
                 Phys. Lett. B 412 (1997) 95;
                 N.G. Stefanis, in
                 XI International Conference on Problems of Quantum
                 Field Theory in memory of D.I. Blokhintsev,
                 July 13-17, 1998, JINR, Dubna, Russia,
                 hep-ph/9811262.

\bibitem{LB80} G.P. Lepage and S.J. Brodsky,
               Phys. Rev. D 22 (1980) 2157.

\bibitem{ER80} A. V. Efremov and A. V. Radyushkin,
               Riv. Nuovo Cimento {\bf 3} (1980) 1.

\bibitem{CS81} J.C. Collins and D.S. Soper,
               Nucl. Phys. B 193 (1981) 381;
               B 194 (1982) 445;
               J.C. Collins, D.E. Soper, and G. Sterman,
               Nucl. Phys. B 250 (1985) 199.

\bibitem{CZ84} V.L. Chernyak and A.R. Zhitnitsky,
               Phys. Rep. 112 (1984) 173.

\bibitem{ILS84} N. Isgur and C.H. Llewellyn-Smith,
                Phys. Rev. Lett. 52 (1984) 1080;
                Phys. Lett. B 217 (1989) 535;
                Nucl. Phys. B 317 (1989) 526.

\bibitem{BS89} J. Botts and G. Sterman,
               Nucl. Phys. B 325 (1989) 62.

\bibitem{CSS89} J. C. Collins, D. E. Soper, and Sterman, in
                Perturbative Quantum Chromodynamics,
                ed. A. H. Mueller
                (World Scientific, Singapore, 1989),
                p. 1.

\bibitem{LS92} H.-n. Li and G. Sterman,
               Nucl. Phys. B 381 (1992) 129.
               H.-N. Li,
               Phys. Rev. D 48 (1993) 4243.

\bibitem{BJKBS95} J. Bolz, R. Jakob, P. Kroll,
                  M. Bergmann, and N.G. Stefanis,
                  Z. Phys. C 66 (1995) 267;
                  N.G. Stefanis,
                  Mod. Phys. Lett. A10 (1995) 1419.

\bibitem{Szc96} A. Szczepaniak,
                Phys. Rev. D 54 (1996) 1167;
                A. Szczepaniak, A. Radyushkin, and C.-R. Ji,
                Phys. Rev. D 57 (1998) 2813.

\bibitem{SSK98} N.G. Stefanis, W. Schroers, and H.-Ch. Kim,
                Phys. Lett. B 449 (1999) 299;
                hep-ph/9812280.

\bibitem{SS94} M.G. Sotiropoulos and G. Sterman,
               Nucl. Phys. B 419 (1994) 59.


\bibitem{KOS98} N. Kidonakis, G. Oderda, and G. Sterman,
                Nucl. Phys. B 531 (1998) 365.

\bibitem{Li97} H.-n. Li,
               Phys. Rev. D 55 (1997) 105.

\bibitem{Li88} L.N. Lipatov,
               Nucl. Phys. B 309 (1988) 379.

\bibitem{Kor94} G.P. Korchemsky,
                Phys. Lett. B 325 (1994) 459;
                I.A. Korchemskaya and G.P. Korchemsky,
                Nucl. Phys. B 437 (1994) 127.

\bibitem{KR86} G.P. Korchemsky and A.V. Radyushkin,
               Phys. Lett. B 171 (1986) 459;
               Yad. Fiz. 45 (1987) 1466
               [Sov. J. Nucl. Phys. 45 (1987) 910];
               Nucl. Phys. B 283 (1987) 342.

\bibitem{Bra81} R.A. Brandt, F. Neri, and M.-A. Sato,
                Phys. Rev. D 24 (1981) 879.

\bibitem{Sen81} A. Sen,
                Phys. Rev. D 24 (1981) 3281.
\end{thebibliography}
\end{document}